# Harrison–Zel'dovich scale invariance and the exponential decrease of the "cosmological constant" in the super-early Universe

O V Babourova[1] and B N Frolov[2]

[1] Department 'Physics', Moscow Automobile and Road Construction State Technical University (MADI), Leningradsky pr., 64, Moscow, 125319, Russia
[2] Institute of Physics, Technology and Information Systems, Moscow Pedagogical State University (MSPU), M. Pirogovskaya str. 29/7, Moscow 119435, Russia

E-mail: ovbaburova@madi.ru, bn.frolov@mpgu.su

**Abstract.** Cosmological consequences of the Poincaré–Weyl gauge theory of gravity are considered. A generalized cosmological constant depending from the Dirac scalar field is introduced. The stage of a super-early (Harrison–Zel'dovich) scale invariant Universe is considered. It is shown that while the scale factor sharply increases and demonstrates inflationary behavior, the generalized cosmological constant decreases sharply from a huge value at the beginning of the Big Bang to an extremely small value in the modern era, which solves the well-known "cosmological constant problem".

## 1. Introduction

At the super-early stage of the Universe, when the rest masses of elementary particles were not yet emerged, all interactions were carried out by massless quanta. These interactions have the property of scale invariance, as it has been suggested by Zel'dovich and Harrison. The scale invariance hypothesis underlies the calculation of the initial part of the spectrum of the primary fluctuations of the density of matter in the early Universe (Harrison – Zel'dovich plateau, see Figure 1 and Figure 2).

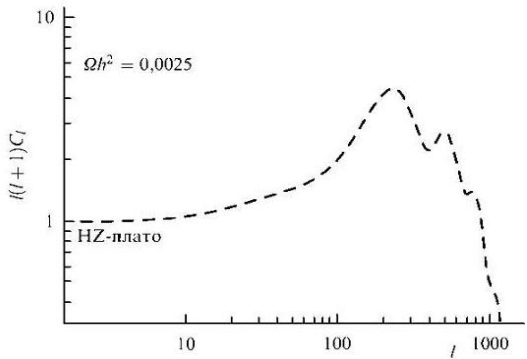 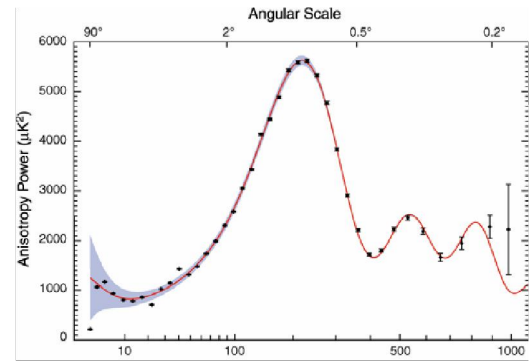

**Figure 1.** The Harrison–Zel'dovich Plateau at small $l$ [1].

**Figure 2.** The results of the COBE experiment for the study of the anisotropy of the brightness of the background radiation [1].

In this case, the structure of the tangent space of the space-time manifold in the super-early Universe is not determined by the Poincaré group, but by the Poincaré–Weyl group, in which the transformations of the Poincaré group are complemented by the transformations of the Weyl subgroup − extensions and contractions (dilatations) of spacetime. In [2 − 4] the gauge theory of the Poincaré–Weyl group was developed. It was shown that the space-time is Cartan–Weyl space with the curvature 2-form $R^a{}_b$ and the torsion 2-form $T^a$, as well as the nonmetricity 1-form $Q_{ab}$ with the Weyl condition $Q_{ab} = (1/4) g_{ab} Q$. An addition to the metric tensor, an additional Weyl scalar field $\beta$ appears, which coincides with the scalar field introduced by Dirac [5].

## 2. Lagrangian density and field equations

In the Poincaré–Weyl gauge theory of gravity, the 4-form of Lagrangian density has the form (here the terms with squares of curvature are omitted),

$$L = 2f_0\beta^2 \left[ (1/2) R^a{}_b \wedge \eta_a{}^b - \beta^2 \Lambda_0 \eta + \rho_1 T^a \wedge *T_a + \rho_2 (T^a \wedge \theta_b) \wedge *(T^b \wedge \theta_a) \right.$$
$$+ \rho_3 (T^a \wedge \theta_a) \wedge *(T^b \wedge \theta_b) + \xi Q \wedge *Q + \zeta Q \wedge \theta^a \wedge *T_a + l_1 \beta^{-2} d\beta \wedge *d\beta \quad (1)$$
$$\left. + l_2 \beta^{-1} d\beta \wedge \theta^\alpha \wedge *T_\alpha + l_3 \beta^{-1} d\beta \wedge *Q \right] + \beta^4 \Lambda^{ab} \wedge (Q_{ab} - (1/4) g_{ab} Q).$$

Here $*$ is the Hodge dualization operation, $\eta$ is a volume 4-form with components $\eta_{abcd}$, $\eta_{ab} = (1/2!) \theta^c \wedge \theta^d \eta_{abcd}$ is 2-form, $\Lambda^{ab}$ are Lagrange multipliers.

The term $\beta^2 \Lambda_0$ describes the effective cosmological constant (dark energy, the energy of the physical vacuum, see E.B. Gliner [6, 7]), where $\Lambda_0$ is the theory parameter providing the correct rate of inflation. The equality $\beta_0^2 \Lambda_0 = 10^{120} \Lambda$ (see [8]) should be fulfilled, where $\Lambda$ is the modern value of the cosmological constant, and $\beta_0$ is the value of the Dirac scalar field at $t = 0$.

The technique of variation calculus in the language of external forms in the Cartan–Weyl space was developed on the basis of the lemma on commutation of variation and the dualization operation [9, 10], and then modified to take into account the introduction of the Dirac scalar field $\beta$:

As a result, we have three field equations: Γ-equation, $\theta$-equation, and $\beta$-equation.

Γ-equation:

$$f_0 \left[ -(1/4) Q \wedge \eta_a{}^b + (1/2) T_c \wedge \eta_a{}^{bc} + (1/2) \eta_{ac} \wedge Q^{bc} + d\ln\beta \wedge \eta_a{}^b + \right.$$
$$+ 2\rho_1 \theta^b \wedge *T_a + 2\rho_2 \theta^b \wedge \theta_c \wedge *(T^c \wedge \theta_a) + 2\rho_3 \theta^b \wedge \theta_a \wedge *(T^c \wedge \theta_c) +$$
$$+ 4\xi \delta_a^b [2, 3, 4] Q + \zeta (2\delta_a^b \theta^c \wedge *T_c + \theta^b \wedge *(Q \wedge \theta_a)) + \quad (2)$$
$$\left. + l_2 \theta^b \wedge *(d\ln\beta \wedge \theta_a) + l_3 2\delta_a^b * d\ln\beta \right] - \beta^2 \Lambda_a{}^b = 0.$$

$\theta$-equation:

$$\frac{1}{2} R^b{}_c \wedge \eta_{ba}{}^c + \rho_1 \left[ 2D*T_a + T_b \wedge *(T^b \wedge \theta_a) + *(T^b \wedge \theta_a) \wedge *T_b + 4d\ln\beta \wedge *T_a \right] +$$
$$+ \rho_2 \left[ 2D(\theta_b \wedge *(T^b \wedge \theta_a)) + 2T^b \wedge *(\theta_b \wedge T_a) - *(T^b \wedge \theta_c \wedge \theta_a)(T^c \wedge \theta_b) \right.$$
$$\left. - *(*(T^c \wedge \theta_d) \wedge \theta_a) \wedge *(T^d \wedge \theta_c) + 4d\ln\beta \wedge \theta_b \wedge *(T^b \wedge \theta_a) \right] +$$
$$+ \rho_3 \left[ 2D(\theta_a \wedge *(T^b \wedge \theta_b)) + 2T_a \wedge *(T^b \wedge \theta_b) - *(T^b \wedge \theta_b \wedge \theta_a)(T^c \wedge \theta_c) \right.$$
$$\left. - *(*(T^b \wedge \theta_b) \wedge \theta_a) \wedge *(T^c \wedge \theta_c) + 4d\ln\beta \wedge \theta_a \wedge *(T^b \wedge \theta_b) \right] +$$
$$+ \xi \left[ -Q \wedge *(Q \wedge \theta_a) - *(*Q \wedge \theta_a) *Q \right] +$$
$$+ \zeta [D*(Q \wedge \theta_a) - Q \wedge *T_a + Q \wedge \theta^b \wedge *(T_b \wedge \theta_a) +$$
$$+ *(*T_b \wedge \theta_a) \wedge *(Q \wedge \theta^b) + 2d\ln\beta \wedge *(Q \wedge \theta_a)] +$$
$$+ l_1 [-d\ln\beta \wedge *(d\ln\beta \wedge \theta_a) - *(*d\ln\beta \wedge \theta_a) \wedge *d\ln\beta)] + \quad (3)$$
$$+ l_2 [D*(d\ln\beta \wedge \theta_a) + d\ln\beta \wedge \theta^b \wedge *(T_b \wedge \theta_a) - d\ln\beta \wedge *T_a +$$
$$+ *(*T_b \wedge \theta_a) \wedge *(d\ln\beta \wedge \theta^b) + 2d\ln\beta \wedge *(d\ln\beta \wedge \theta_a)] +$$
$$+ l_3 [-d\ln\beta \wedge *(Q \wedge \theta_a) - *(*Q \wedge \theta_a) *d\ln\beta] = 0.$$

Here D is a tensor-value covariant external differential,

$$D\Psi^a{}_b = d\Psi^a{}_b + \Gamma^a{}_c \wedge \Psi^c{}_b - \Gamma^c{}_b \wedge \Psi^a{}_c.$$

$\beta$-equation:

$$R^a{}_b \wedge \eta_a{}^b + \rho_1 2T^a \wedge *T_a + \rho_2 2(T^a \wedge \theta_b) \wedge *(T^b \wedge \theta_a) +$$
$$+ \rho_3 2(T^a \wedge \theta_a) \wedge *(T^b \wedge \theta_b) + \xi 2Q \wedge *Q + \zeta 2Q \wedge \theta^a \wedge *T_a + \quad (4)$$
$$+ l_1(-2d*d\ln\beta - 2d\ln\beta \wedge d\ln\beta) + l_2(-d(\theta^a \wedge *T_a)) + l_3(-d*Q) = 0.$$

## 3. Cosmological model

Let us apply the Poincaré–Weyl theory of gravity [2 − 4] to the super-early stage of a homogeneous isotropic and spatially flat Universe with the Friedman–Robertson–Walker metrics,

$$ds^2 = dt^2 - a^2(t)(dx^2 + dy^2 + dz^2).$$

For a given metric, the 2-form of torsion $T_a = (1/3)T \wedge \theta_a$ is completely determined by the 1-form of its trace $T = *(\theta_c \wedge *T^a)$. In this case the antisymmetric part of the Γ-equation (2) is equivalent to the equations,

$$2(\rho_1 - 2\rho_2 - 1)T + 3(1/4 + \zeta)Q \equiv (6 - 3l_2)(d\beta/\beta), \quad (5)$$

$$2(\rho_1 - 2\rho_2 + 4\zeta)T + (16\xi + 3\zeta)Q \equiv (-3l_2 - 8l_3)(d\beta/\beta). \quad (6)$$

The symmetric part of the Γ-equation defines the 3-form of the Lagrange multipliers $\Lambda^{ab}$.

As a result, the 1-form of the torsion trace T and the 1-form of the nonmetricity trace Q are determined by the field $\beta$,

$$T = s(d\beta/\beta), \quad Q = q(d\beta/\beta),$$

where $s$, $q$ are the numbers determined by the coupling constants of the Lagrangian density (1)

Let us introduce the notations ($a_t = da/dt$, $\beta_t = d\beta/dt$, H is the Hubble constant),

$$H = a_t/a, \quad U = \beta_t/\beta, \quad k^2 = l_1 + (1/2)sl_2 + (1/2)ql_3.$$

We accept the condition, $q/8 - s/3 = 1$, which was also accepted when finding spherically and axially symmetric solutions of the of Poincaré–Weyl theory of gravity [11−13].

Then the field equations take the form,

$$3H^2 + 6HU + U^2(k^2 + 3) = \Lambda_0 \beta^2, \quad (7)$$

$$2H_t + 2U_t + 4HU + 3H^2 - U^2(k^2 - 1) = \Lambda_0 \beta^2, \quad (8)$$

$$3H_t + 6H^2 + (U_t + U^2 + 3HU)(k^2 + 3) = 2\Lambda_0 \beta^2. \quad (9)$$

Here the first two equations are the $\theta$-equations, and the third equation is the $\beta$-equation.

Subtracting equation (7) from equation (8), we get

$$H_t + U_t - HU - U^2(k^2 + 1) = 0. \quad (10)$$

Subtracting from equation (9) the double equation (8) and taking into account the equation (10), we obtain the equation

$$k^2(U_t + 3HU + 2U^2) = 0, \quad k^2 \neq 0. \quad (11)$$

When $k^2 \neq 0$, we obtain for two unknown functions $a(t)$ and $\beta(t)$ three equations: (7), (10) and (11). Thus, the system of equations is over-determined.

We estimate the value of $k^2$. Based on the calculation of the flyby anomaly (upon receipt of the spherically symmetric solution [11, 12]), the following estimate was obtained,

$$k^2 \approx 10^{-20} \div 10^{-22}. \quad (12)$$

Therefore, in equations (7) and (10), we may neglect the terms with the coefficient $k^2$.

Then equation (7) will be reduced to the equation

$$H + U = (\lambda/3)\beta, \quad \lambda = \sqrt{3\Lambda_0}. \quad (13)$$

Here from the two possible signs, we take the "+" sign.

Equation (10) with regard to (12) turns out to be a consequence of equation (13), and the system of equations reduces to the two equations (11) and (13) for two unknown functions $a(t)$ and $\beta(t)$, and the system of equations is becoming well defined.

Due to the indicated property of the system of equations, the system of equations (7), (10) and (11) with $k^2 = 0$ turns out to be underdetermined, because in this case there is only one equation (13) for two unknown functions.

Equation (11) with regard to (13) is reduced to

$$U_t + \lambda\beta U - U^2 = 0. \tag{14}$$

This equation is equivalent to the equation

$$\beta\beta_{tt} + \lambda\beta_t\beta^2 - \beta_t^2 = 0, \tag{15}$$

which is integrated by substitution $\beta_t = \beta^2 z(\beta)$. The first integral of this equation is

$$U = \beta_t/\beta = -\lambda\beta\ln(C\beta), \tag{16}$$

where $C$ is an integration constant.

Substituting this solution into equation (13), we obtain the second equation of the system,

$$H = a_t/a = \lambda\beta\left(\ln(C\beta) + 1/3\right). \tag{17}$$

We put a boundary condition,

$$C\beta \to 1 \quad \text{when} \quad t \to \infty. \tag{18}$$

One can obtain from the equations (15) and (16) that the approach to the limit value $C\beta \to 1$ at large $t$ occurs exponentially.

The value of the constant $C$ is obtained from the requirement that the value of the effective cosmological constant $\beta^2(t)\Lambda_0$ coincides with the modern value of the cosmological constant,

$$\beta^2(t_U)\Lambda_0 = \Lambda, \qquad t_U = 13.8 \text{ billion years}, \tag{19}$$

where $t_U$ is the lifetime of the Universe.

We produce the numerical integration of the system of equations (16) and (17) with the boundary condition (18). Figures 3 and 4 shows the solution models for small and large values of time $t$. We see that the cosmological solution obtained demonstrates the inflationary behavior for the scale factor $a(t)$, while the generalized cosmological constant $\beta^2(t)\Lambda_0$ decreases sharply from a huge value at the beginning of Big Bang to an extremely small (but not zero) value in the modern era, which coincides with its observable value.

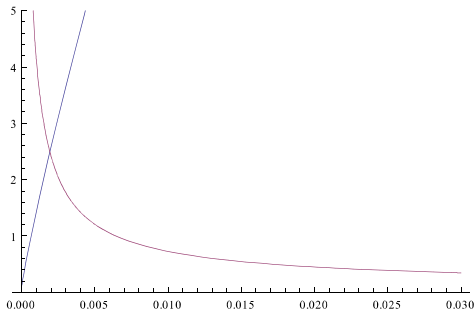 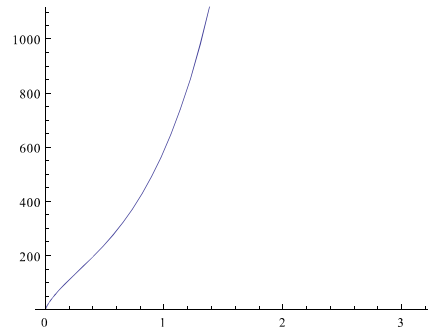

**Figure 3**. Model of behavior of the functions $a(t)$, $\beta(t)$ for small values of time $t$.

**Figure 4**. Model of behavior of the functions $a(t)$, $\beta(t)$ for large values of time $t$.

## 4. Conclusion

The hypothesis on the sharply decrease of the effective cosmological constant as a consequence of the fields dynamics in the super-early Universe was expressed in the monograph [10] and the papers [14−16]. The main thing of the solution obtained and that found earlier by the authors is that these solutions demonstrates for the effective cosmological constant sharply decrease at small $t$ and approach $\Lambda$ (but not zero) at large values of $t$.

The result obtained can be considered as a solution to one of the main contradictions of the theory of the evolution of the Universe, which is called the "problem of the cosmological constant" [8, 17−19]. The essence of this problem is a huge (about 120 order) difference between the value of the physical vacuum energy, described by the cosmological constant, in the initial stage of the Universe evolution (determined on the basis of quantum field theory) and its value, determined on the basis of modern observational data. The problem of the cosmological constant is one of the important problems of modern fundamental physics [19]. The solution of this problem will allow to reconcile the theory of the evolution of the Universe with modern physical concepts.

The main assumption, which was used to obtain the main result of this article, is the Harrison−Zel'dovich hypothesis about the scale invariance of the super-early Universe, confirmed by the observations of the temperature inhomogeneity of the CMB. A corollary of this hypothesis is the fact that the structure of the tangent space of the space-time manifold of the super-early Universe is determined not by the Poincaré group as the fundamental group, but by the Poincaré –Weyl group, in which the transformations of the Poincaré group are supplemented by the Weyl subgroup of scale transformations, and this scale invariance is slightly violated.

The notion of weakly violated scale invariance of the super-early Universe inevitably entails a radical change in ideas about the properties of space-time, namely, the recognition of the fact that the fundamental group of space-time in the modern era is also the Poincaré −Weyl group (neglecting the effects of gravitation), but already with strongly broken scale symmetry (due to the birth of particles with nonzero rest mass). This circumstance can serve as an explanation of the existence of the nonlocality phenomenon (action at a distance), recently discovered in quantum physics, as well as in some macroscopic phenomena (see [20, 21]).